\documentclass{iau}
\usepackage{graphicx}

\title[Spectroscopic and photometric study of HD\,101794 and HD\,167003] %% give here short title %%
{Spectroscopic and photometric study of two B-type pulsators in eclipsing systems}

\author[Dominik Drobek \& Andrzej Pigulski]   %% give here short author list %%
{Dominik Drobek
 \and Andrzej Pigulski}

\affiliation{Instytut Astronomiczny, University of Wroc{\l}aw, \\ 
ul. Kopernika~11, 51-622 Wroc{\l}aw, Poland\\
email: {\tt drobek@astro.uni.wroc.pl}}

\pubyear{2013}
\volume{301}  %% insert here IAU Symposium No.
\pagerange{1--2}
\jname{Precision Asteroseismology}
\editors{J.A. Guzik, W.J. Chaplin, G. Handler \& A. Pigulski, eds.}
\begin{document}

\maketitle

\begin{abstract}
Pulsating stars in eclipsing binary systems play an important role in asteroseismology. The combination of their spectroscopic and
photometric orbital solutions can be used to determine, or at least to constrain, the masses and radii of components. To
successfully perform any seismic modelling of a star, one has to identify at least some of the detected modes, which requires
precise time-series photometric and spectroscopic observations. This work presents a progress report on the analysis of two 
$\beta$ Cephei-type stars in eclipsing binaries: HD\,101794 (V916 Cen) and HD\,167003 (V4386 Sgr).
\keywords{stars: individual (HD\,101794, HD\,167003) -- stars: oscillations -- binaries: eclipsing}
\end{abstract}

\firstsection % if your document starts with a section,
\section{Observations and data reduction}
HD\,101794 and HD\,167003 have been observed at the South African Astronomical Observatory (SAAO) between May 2 and 19, 2009. The
$UBVRI$ time-series photometry has been acquired using the UCT CCD detector at the 1.0-m telescope. Spectroscopic observations were
carried out using the GIRAFFE {\'e}chelle spectrograph at the 1.9-m telescope. We obtained spectra in the wavelength range between 
4200 {\AA} and 6900 {\AA} with a resolution $R \approx$ 39000. Stellar magnitudes were calculated using the program DAOPHOT II
(\cite[Stetson 1987]{stet87}). Spectra were wavelength-calibrated and extracted using the IRAF software package 
(\cite[Tody 1993]{tody93}). The {\'e}chelle orders were merged and normalised with a program of our own, and the radial
velocities were calculated by cross-correlating the observed spectra with non-LTE models of \cite[Lanz \&~Hubeny (2007)]{lahu07}
using the method of \cite[Tonry \&~Davis (1979)]{toda79}.

\section{Data analysis and results}
The spectrum of HD\,101794 features very broad lines, the broadening being caused by rapid rotation. This is not surprising at
all, since HD\,101794 is a known Be star. Unfortunately, the resulting broadening reduces the accuracy of the radial velocities
obtained with cross-correlation. For this reason the analysis of HD\,101794 requires more consideration. The star is still under
study, and the results will not be discussed here.

The radial velocity time-series data of HD\,167003 were subjected to Fourier analysis. The orbital period was estimated at
10.88~d, which is close to the value of 10.79824~d obtained from the analysis of the All Sky Automated Survey phase 3 (ASAS-3)
$V$-band photometry by \cite[Pigulski \&~Pojma{\'n}ski (2008)]{pipo08}. We adopted their period value in our subsequent analysis.
Apart from the effects of orbital motion, at least four frequencies arising from  stellar pulsations are present in the power
spectrum. The frequency $f_1$ = 7.351~d$^{-1}$ is only seen in the radial velocity  data. Our $f_2$ = 6.771~d$^{-1}$ and 
$f_3$ = 7.023~d$^{-1}$ correspond to $f_1$ and $f_3$ found by \cite[Pigulski \&~Pojma{\'n}ski (2008)]{pipo08}. Our last frequency,
$f_4$ = 8.451~d$^{-1}$, is not seen in their photometric data, while their $f_4$ was not detected in our radial velocity measurements. In our
newly acquired SAAO multicolour photometry, we detected frequencies corresponding to $f_1$, $f_2$ and $f_3$ found by the
aforementioned authors.

Once the frequencies of pulsation were determined, their contribution was removed from the original radial velocity data. We
attempted to model the radial velocity changes arising from orbital motion, and arrived at the following set of parameters:
orbital period $P_{\rm{orb}}$ = 10.79824~d (fixed); semi-amplitude $K$ = (31.8 $\pm$ 0.6) km/s; eccentricity $e$ = 0.061 $\pm$ 
0.013; argument of periastron $\omega$ = (299 $\pm$ 8)$^{\circ}$; systemic velocity $\gamma$ = ($-$30 $\pm$ 0.4) km/s and the
time of periastron passage $T_0$ = HJD 2454965.53 $\pm$ 0.23. The standard deviation from the fit amounts to 2.9 km/s.

\section{Discussion}
The results for HD\,167003 are very encouraging. First of all, we have confirmed it is indeed a pulsating star in a binary system.
In addition, we have confirmed that the orbital period amounts to 10.79824 d and is not twice as long, as initially suspected by
\cite[Pigulski \&~Pojma{\'n}ski (2008)]{pipo08}. This star seems to be a single-lined spectroscopic binary, and our modelling
suggests that the orbit is close to circular.

\cite[Pigulski \&~Pojma{\'n}ski (2008)]{pipo07} detected only the primary eclipse in the ASAS-3 photometry of HD\,167003. Our 
initial hypothesis was that the lack of the secondary eclipse is caused by a highly eccentric orbit. In light of the results of 
our modelling, we now know this cannot be the case. This suggests that the secondary eclipse is very shallow, and that the 
contribution of the secondary component to the total flux is small. While our present photometry is more accurate than the ASAS-3
photometry used by the previous investigators, we are also unable to find the secondary minimum in our observations. This could 
be because the orbital period of HD\,167003 is quite long, and our phase coverage is incomplete. However, we managed to detect
hints of a minute reflection effect.

Once the orbital inclination is known from the light curve modelling, we will be able to use the mass function to constrain the
masses of components. From the fact that at least two of the photometrically detected modes are also seen in the radial velocity
data, it seems probable that the attempts to identify mode degrees from amplitude ratios and phase differences will be 
successful. Therefore, this star seems to be a very good candidate for asteroseismic analysis.

\begin{acknowledgements}
The authors would like to thank G.~Kopacki for acquiring photometric observations of our targets. D.~Drobek would like to thank
E.~Niemczura for her help with the model spectra, and Z.~Ko{\l}aczkowski and J.~Molenda-{\.Z}akowicz for their assistance with
the IRAF software package. This work has been supported by the National Science Centre grants no. 2011/01/N/ST9/00400 and
no. 2011/03/B/ST9/02667.
\end{acknowledgements}

\end{document}